\documentclass{my_new_tlp}
\pdfoutput=1

\usepackage{tikz} 
\usetikzlibrary{shapes,arrows} 

\definecolor{lightblue}{rgb}{0.145,0.6666,1} 

\usepackage{amssymb} 

\usepackage{hyperref}

\newcommand{\bee}{\textsf{BEE}}
\newcommand{\nauty}{\texttt{nauty}}

 \newcommand{\gtools}{\texttt{gtools}}
 \newcommand{\plnauty}{\texttt{pl-nauty}}
 \newcommand{\plgtools}{\texttt{pl-gtools}}

 \newcommand{\comment}[1]{}

 \newcommand{\set}[1]{\left\{
       \begin{array}{l}#1\end{array}
       \right\}}

 \newcommand{\RR}{{\cal R}}

 \title[Logic Programming with Graph Automorphism]
       {Logic Programming with Graph Automorphism: Integrating
        \nauty~with Prolog (Tool Description)\thanks{Supported by the 
              Israel Science Foundation, grant 182/13.}}
 \author[Michael Frank and Michael Codish]{Michael Frank and Michael Codish \\
         Department of Computer Science\\Ben-Gurion University of the Negev\\
         Beer-Sheva, Israel\\
         \email{\{frankm,mcodish\}@cs.bgu.ac.il}}
 \date{}

\begin{document}

 \maketitle

 \begin{abstract}

   This paper presents the \plnauty~library, a Prolog interface to the
   \nauty~graph-automorphism tool.
   Adding the capabilities of \nauty~to Prolog combines the strength of
   the ``generate and prune'' approach that is commonly used in logic
   programming and constraint solving, with the ability to reduce
   symmetries while reasoning over graph objects. Moreover, it enables
   the integration of \nauty~in existing tool-chains, such as
   SAT-solvers or finite domain constraints compilers which exist for
   Prolog.
   The implementation consists of two components: \plnauty, an
   interface connecting \nauty's C library with Prolog, and
   \plgtools, a Prolog framework integrating the software 
   component of \nauty, called \gtools, with Prolog.
   The complete tool is available as a SWI-Prolog module.  We provide a
   series of usage examples including two that apply to generate Ramsey
   graphs.
   This paper is under consideration for publication in TPLP.

 \end{abstract}

 \section{Introduction}

 Many problems, particularly in combinatorics, reduce to asking whether
 some graph with a given property exists, or alternatively, asking how
 many such non-isomorphic graphs exist.  Such graph search and graph
 enumeration problems are notoriously difficult, in no small part due
 to the extremely large number of symmetries in graphs. In practical
 problem solving, it is often advantageous to eliminate these
 symmetries which arise naturally due to graph isomorphism: typically,
 if a graph $G$ is a solution then so is any other graph $G'$ that is
 isomorphic to $G$.

 General approaches to graph search problems typically involve either:
 \emph{generate and test}, explicitly enumerating all (non-isomorphic)
 graphs and checking each for the given property, or \emph{constrain
   and generate}, encoding the problem for some general-purpose
 discrete satisfiability solver (i.e. SAT, integer programming,
 constraint programming), which does the enumeration implicitly.
 In the explicit approach, one typically iterates, repeatedly applying
 an extend and reduce approach: First \emph{extend} the set of all
 non-isomorphic graphs with $n$ vertices, in all possible ways, to
 graphs with $n+1$ vertices; and then \emph{reduce} the extensions to
 their non-isomorphic (canonical) representatives.
 In the constraint based approach, one typically first encodes the
 problem and then applies a constraint solver in order to produce
 solutions. The (unknown) graph is represented in terms of Boolean
 variables describing it as an adjacency matrix $A$. The encoding is a
 conjunction of constraints that constitute a model, $\varphi_A$, such
 that any satisfying assignment to $\varphi_A$ is a solution to the
 graph search problem.  Typically, symmetry breaking
 constraints~\cite{Crawford96,CodishMPS13} are added to the model to
 reduce the number of isomorphic solutions, while maintaining the
 correctness of the model.

 It remains unknown whether a polynomial time algorithm exists to
 decide the graph isomorphism problem.  
 Nevertheless,
 finding good graph isomorphism algorithms is critical when exploring
 graph search and enumeration problems.  Recently an algorithm was
 published by \citeN{Babai15} which runs in time $O\left( {\exp
     \left( log^c(n) \right) } \right)$, for some constant $c>1$, and 
 solves the graph isomorphism problem. Nevertheless, top of the line 
 graph isomorphism tools use different methods, which are, in practice, 
 faster.


 \citeN{nauty} introduces an algorithm for graph canonization, and its
 implementation, called \nauty\ (which stands for \emph{no
   automorphisms, yes?}), is described in \cite{nauty_impl}.
 In contrast to earlier works, where the canonical representation of a
 graph was typically defined to be the smallest graph isomorphic to it
 (in the lexicographic order), \nauty\ introduced a notion which takes
 structural properties of the graph into account. For details on how
 \nauty~defines canonicity and for the inner workings of the \nauty\
 algorithm see~\cite{nauty,nauty_impl,hartke_nauty,nautyII}.
 In recent years \nauty~has gained a great deal of popularity and
 success. Other, similar tools, are \textsf{bliss}~\cite{bliss} and
 \textsf{saucy}~\cite{saucy}.

 The \nauty\ graph automorphism tool consists of two main components.
 (1) a C library, \nauty, which may be linked to at runtime, that contains
 functions applicable to find the canonical labeling of a graph, and
 (2) a collection of applications, \gtools, that implement an
 assortment of common tasks that \nauty\ is typically applied to. 
 When downloading the tool both components are included. During
 compilation static library files are created for the C library. These
 files may be linked to at runtime, and header files are provided which 
 may be included in foreign C code.
 During compilation, the applications of \gtools\ are compiled into a
 set of command line applications.

 This paper presents a lightweight Prolog interface to both components
 of \nauty\ which we term \plnauty\ and \plgtools.
 The implementation of \plnauty\ is by direct use of Prolog's foreign
 language interface. The implementation of \plgtools\ is slightly more
 complex.  Each \gtools\ application is run as a child process with the
 input and output controlled via unix pipes. The \plgtools\ framework
 provides a set of general predicates to support this type of
 application integration.


 The integration of \nauty\ into Prolog facilitates programming with
 the strengths of the two paradigms: logic programming for solving
 graph search problems on the one hand, and efficient pruning of
 (intermediate) solutions modulo graph isomorphism, on the other.
 It  enables Prolog programs which address graph search
 problems to apply \nauty~natively, through Prolog, in the process of
 graph search and enumeration. Graphs may be generated
 non-deterministically and may be canonized deterministically.
 It also facilitates the interaction with various graph
 representations: those used in \nauty, and those more natural for use
 with Prolog.

 The interface for \nauty\ from within Prolog combines well also with
 other tools and techniques typically applied when addressing graph
 search problems, such as constraint and SAT based programming.
 For example, recent work~\cite{Codish2016}, presents a computer-based
 proof that the Ramsey number $R(4,3,3)=30$, thus closing a long open
 problem concerning the value of $R(4,3,3)$. That paper made extensive
 use of SAT solvers, symmetry breaking techniques, and the \nauty\
 library. It was this experience that led us to implement \plnauty.




 The remaining sections of this paper are organized in the 
 following manner: Section~(\ref{sec:prelim}) introduces the
 definitions used throughout the paper, as well as the running
 example of Ramsey graphs. Section~(\ref{sec:pln}) introduces
 the core of the \plnauty\ library by examples. 
 Section~(\ref{sec:plg}) details the \plgtools\ framework, 
 and details the template used to integrate \gtools\ applications 
 with Prolog. Section~(\ref{sec:tech}) closes some technical
 loose ends, including details of supported platforms, 
 package availability, and additional references to source code.
 Finally, Section~\ref{sec:conclude} concludes.

 \section{Preliminaries}
 \label{sec:prelim}

 A graph $G=(V,E)$ consists of a set of vertices
 $V=[n]=\set{1,\ldots,n}$ and a set of edges $E\subseteq V \times
 V$. In the examples presented in this paper graphs are always
 simple. Meaning that they are undirected, there are no self loops, and
 no multiple edges.
 The tools we present allow also directed graphs and support
 vertex-coloring. 
 %

 Two graphs $G=([n],E)$ and $G'=([n],E')$ are said to be isomorphic if
 the vertices of one graph may be permuted to obtain the other. Namely,
 if there exists a permutation $\pi\colon[n]\to[n]$ such that $(u,v)
 \in E \iff (\pi(u),\pi(v)) \in E'$. 
 %
 %
 Graph isomorphism is an equivalence relation. As such, it induces
 equivalence classes on any set of graphs, wherein graphs $G$, $G'$ are
 in the same equivalence class if $G$ and $G'$ are isomorphic. 
 The canonical representation of a graph $G$ is some fixed value $can(G)$
 such that for every graph $G'$ isomorphic to $G$ we have $can(G) =
 can(G')$.

 The running example we use throughout this paper concerns the generation
 of Ramsey graphs: A $R(s,t;n)$ Ramsey graph, where $s,t,n\in\mathbb{N}$,
 is a graph $G$ with $n$ vertices such that $G$ contains no clique of size 
 $s$ nor an independent set of size $t$. We denote by $\RR(s,t;n)$ the set 
 of all non-isomorphic Ramsey $R(s,t;n)$ graphs. The Ramsey number $R(s,t)$ 
 is the smallest natural number $n$ for which no $R(s,t;n)$ graph exist.

 \section{Interfacing Prolog with \nauty's C library}
 \label{sec:pln}

 The \plnauty\ interface is implemented using the foreign language
 interface of SWI-Prolog~\cite{swipl}. The \nauty\ C library is linked
 with corresponding C code written for Prolog, which involves four
 low-level Prolog predicates: 
 (1)~\texttt{densenauty/8}, 
 (2)~\texttt{canonic\_graph/6},
 (3)~\texttt{isomorphic\_graphs/6}, and
 (4)~\texttt{graph\_convert/5}.
 The experienced \nauty\ user will find \texttt{densenauty/8} to be a
 direct interface to the corresponding C function in \nauty. The
 \texttt{canonic\_graph/6} predicate performs graph canonization
 only. The \texttt{isomorphic\_graphs/6} predicate tests two graphs for
 isomorphism, and  \texttt{graph\_convert/5} converts between the
 supported graph formats such as between the
 \texttt{graph6}~\cite{graph6} format often used in \nauty\ and the
 Boolean adjacency matrices natural in logic programming.

 We present several examples of the \plnauty\ library in Prolog. The first
 two examples revolve around enumerating Ramsey graphs modulo isomorphism.
 The rest are simple demonstrations of the core \plnauty\ predicates in various 
 cases.
 In the first example we apply a straightforward iterative approach to
 enumerate all solutions modulo isomorphism.  
 The second example illustrates how \nauty~ integrates into an existing
 tool-chain, all specified as part of the Prolog process. Here we first
 construct a constraint model, infused with a partial symmetry breaking
 predicate. Then, apply the finite domain constraint compiler
 \bee~\cite{BEE,jair2013} (written in Prolog) to obtain a CNF model, apply a SAT
 solver (through its Prolog interface), and then generate all solutions
 of constraint model. At the end of each iteration we apply predicates 
 from the \plnauty\ library to remove isomorphic solutions.
 The core of the code, with an emphasis on using the \plnauty~library
 is presented below. The complete code is available for download as
 part of the \plnauty~library, in the \verb|examples| directory.

 \subsection{The First Example: Generate and Test}

 In the code below,
 the predicate \verb!genRamseyGT(S, T, N, Graphs)! iterates starting from
 the empty graph to generate in \verb!Graphs!, the set of all canonical
 Ramsey $(S,T;N)$ colorings. We represent graphs as Boolean adjacency
 matrices: a list of \verb!N! length-\verb!N! lists.
 At iteration \verb!I! it takes, \verb!Acc!, the canonical set of
 Ramsey $(S,T;I)$ colorings computed thus far and calls the predicate
 \verb!extendRamsey(S, T, I, Acc, NewAcc)! to obtain, \verb!NewAcc!, the
 canonical set of Ramsey $(S,T;I+1)$ colorings. 
 {\scriptsize
 \begin{verbatim}
     genRamseyGT(S, T, N, Graphs) :-
         genRamsey(0, S, T, N, [[]], Graphs).

     genRamsey(I, S, T, N, Acc, Graphs) :-
         I < N, !, I1 is I+1,
         extendRamsey(S, T, I, Acc, NewAcc),
         genRamsey(I1, S, T, N, NewAcc, Graphs).
     genRamsey(N, _, _, N, Graphs, Graphs).
 \end{verbatim}
 }

 The predicate \verb!extendRamsey(S, T, N, Graphs, NewGraphs)!  takes a
 list, \verb!Graphs! of (canonical) Ramsey $(S,T;N)$ graphs.  Then, a
 new vertex is added in all possible ways to each graph in
 \verb!Graphs! and those new graphs that are Ramsey $(S,T;N+1)$
 colorings are canonized. Finally, the resulting  graphs  are  sorted
 to remove duplicates, resulting in \verb!NewGraphs!. It is the call to
 \verb!canonic_graph/3! that interfaces to our Prolog integration of
 the \nauty\ tool.
 {\scriptsize
 \begin{verbatim}
     extendRamsey(S, T, N, Graphs, NewGraphs) :-
         N1 is N+1,
         findall(Canonic,
                 (member(Graph, Graphs),
                  addVertex(Graph, NewGraph),
                  isRamsey(S,T,N1,NewGraph),             /* #1 (test)*/
                  canonic_graph(N1, NewGraph, Canonic)   /* #2 (reduce)*/ 
                 ),
                 GraphsTmp),
         sort(GraphsTmp, NewGraphs).
 \end{verbatim}
 }

 The predicate \verb!addVertex(Matrix,ExtendedMatrix)! extends
 non-deterministically an adjacency \verb!Matrix! with  a new vertex by
 adding a new first row and equal first column.
 {\scriptsize
 \begin{verbatim}
     addVertex(Matrix,[NewRow|NewRows]) :-
         NewRow = [0|Xs],
         addFirstCol(Matrix,Xs,NewRows).

     addFirstCol([],[],[]).
     addFirstCol([Row|Rows],[X|Xs],[[X|Row]|NewRows]) :-
         member(X,[0,1]),
         addFirstCol(Rows,Xs,NewRows).
 \end{verbatim}
 }

 To complete the example, we illustrate the test predicate
 \verb!isRamsey(S,T,N,Graph)! which succeeds if the given \verb!Graph!
 is a Ramsey $(S,T;N)$ coloring. This is so if it is not possible to
 \verb!choose! \verb!S! vertices from the graph, the edges between
 which are all ``colored'' 0, or \verb!T! vertices from the graph, the
 edges between which are all ``colored'' 1.
 {\scriptsize
 \begin{verbatim}
       isRamsey(S,T,N,Graph) :-
               forall( choose(N, S, Vs), mono(0, Vs, Graph) ),
               forall( choose(N, T, Vs), mono(1, Vs, Graph) ).

       mono(Color, Vs, Graph) :-
               cliqeEdges(Vs,Graph,Es),
               maplist(==(Color), Es).

       cliqeEdges([],_,[]).                            choose(N,K,C) :-
       cliqeEdges([I|Is],Graph,Es) :-                        numlist(1,N,Ns),
             cliqeEdges(I, Is, Graph, Es0),                  length(C,K),
             cliqeEdges(Is, Graph, Es).                      choose(C,Ns).

       cliqeEdges(_,[],_,[]).                          choose([],[]).
       cliqeEdges(I,[J|Js], Graph, [E|Es]) :-          choose([I|Is],[I|Js]) :-
             nth1(I, Graph, Gi),                             choose(Is,Js).
             nth1(J, Gi, E),                           choose(Is,[_|Js]) :-
             cliqeEdges(I,Js,Graph,Es).                      choose(Is,Js).
 \end{verbatim}
 }

 We first demonstrate the application of the \verb!genRamseyGT! to the so
 called, ``party problem''. What is the smallest number of people that
 must be invited to a party so that at least three know each other, or
 at least three do not know each other. This is the smallest $N$ for
 which there is no $(3,3;N)$ coloring. The following two calls
 illustrate that there is a single canonical coloring when $N=5$ and
 none when $N=6$. So, the answer to the party problem (as well-known)
 is 6.
 {\scriptsize
 \begin{verbatim}
     ?- genRamseyGT(3,3,5,Gs).
     Gs = [ [[0,1,1,0,0],
             [1,0,0,1,0],
             [1,0,0,0,1],
             [0,1,0,0,1],
             [0,0,1,1,0]]].
     ?- genRamseyGT(3,3,6,Gs).
     Gs = [].
 \end{verbatim}
 }

 We make three observations regarding the generation of graphs in this example.
 Consider the predicate \verb!extendRamsey/5!.
 \begin{enumerate}
 \item If the call \verb!canonic_graph(N1, NewGraph, Canonic)!, at the
   line marked \verb!/* #2 */!, is replaced by the line
   \verb!Canonic = NewGraph!, then all solutions are found, not just
   the canonical ones.  For example, when $N=5$ there are 12
   solutions, all of them isomorphic.
 {\scriptsize
 \begin{verbatim}
     ?- genRamseyGT(3,3,5,Gs), length(Gs,M).
     M = 12.
 \end{verbatim}
 }
 \item If the call to \verb!isRamsey(S,T,N1,NewGraph)!, at the
   line marked \verb!/* #1 */!, is removed then
   we generate all non-isomorphic graphs on \verb!N! vertices. For
   example, 
 {\scriptsize
 \begin{verbatim}
     ?- genRamseyGT(3,3,5,Gs), length(Gs,M).
     M = 34.
 \end{verbatim}
 }
 \item If both changes are made, then we generate all graphs on \verb!N! vertices.
 {\scriptsize
 \begin{verbatim}
     ?- genRamseyGT(3,3,5,Gs), length(Gs,M).
     M = 1024.
 \end{verbatim}
 }
 \end{enumerate}

 We now demonstrate the application of the \verb!genRamseyGT! to
 generate incrementally all non-isomorphic $(3,5;N)$ Ramsey
 colorings. It is known \cite{Rad2014} that $R(3,5) = 14$.
 Table~(\ref{tab:35n}) summarizes the enumeration of all non-isomorphic
 $(3,5;N)$ colorings graphs. The first row indicates the number of
 (non-isomorphic) colorings generated. The next rows detail the time
 (in seconds) to compute these colorings and the time spent in the
 calls to \verb!canonic_graph!.  It is notable that the time spent to
 reduce solutions modulo isomorphism using \nauty\ is negligible. 

 \begin{table}[t]
 \centering 
 {\tiny
 \begin{tabular}{c|cccccccccccccc}
   $n$  & 
   $1$  &     
   $2$  &    
   $3$  &   
   $4$  &  
   $5$  & 
   $6$  &
   $7$  &
   $8$  &
   $9$  &
   $10$ &
   $11$ &
   $12$ &
   $13$ &
   $14$ \\ 
 \hline
   $|\RR(3,5;n)|$ 
   & 1          
   & 2             
   & 3             
   & 7             
   & 13            
   & 32            
   & 71            
   & 179           
   & 290           
   & 313           
   & 105           
   & 12            
   & 1             
   & 0  \\     
  time (sec)
  & 0.00 
  & 0.00 
  & 0.00 
  & 0.00 
  & 0.00 
  & 0.03  
  & 0.20   
  & 0.90   
  & 4.66
  & 16.61
  & 39.24
  & 52.72
  & 55.75
  & 56.20   \\
  nauty (sec)
  & 0.00 
  & 0.00 
  & 0.00 
  & 0.00 
  & 0.00 
  & 0.00 
  & 0.01 
  & 0.02 
  & 0.06 
  & 0.09 
  & 0.11 
  & 0.11 
  & 0.11 
  & 0.11 
 \end{tabular}
 }
 \caption{Enumerating $R(3,5;n)$ graphs: Generate, Test \& Reduce.}
 \label{tab:35n}
 \end{table}

 To summarize this section, we stress that this is a toy application
 with the intention to illustrate an application of the integration of
 Prolog with the \nauty\ package.
 A more elaborate solution of this problem would, for example, combine
 the calls
 {\scriptsize
 \begin{verbatim}
    addVertex(Graph, NewGraph), isRamsey(S,T,N1,NewGraph)
 \end{verbatim}
 }
 in \verb!extendRamsey! to add edges connecting the new vertex to the
 rest of the graph incrementally so as not to violate the
 \verb!isRamsey! condition. This combination could also perform various
 propagation based optimizations.

 \subsection{The Second Example: Constrain and Generate}

 In the code below,
 the predicate \verb!genRamseyCG(S, T, N, Graphs)! encodes an instance
 \verb!ramsey(S,T,N)! to a finite domain constraint model. We adopt
 \bee\ ~\cite{BEE,jair2013} for this purpose. The call to \verb!encode/3!
 generates a constraint model, \verb!Constraints! and the
 $\mathtt{N\times N}$ \verb!Matrix! of Boolean (Prolog) variables. The
 \verb!Matrix!  structure serves as a mapping between the instance
 variables, which talks about the search for Ramsey colorings, and the
 \verb!Constraints! variables. It specifies the connection between
 variables in the constraint model and edges in the unknown graph we
 are searching for. The call to \verb!bCompile/2! compiles the
 constraints to a corresponding \verb!CNF!. The call to
 \verb!solveAll/3! iterates with the underlying SAT solver to provide
 all satisfying \verb!Assignments! of the \verb!CNF! (modulo the
 variables of interest in the list \verb!Booleans!). Satisfying
 assignments are then decoded back to the world of graphs in the call
 to \verb!decode/3!, and finally it is here that we call on the
 predicate \verb!canonic_graph/3! from the \plnauty\ interface to
 restrict solutions to their canonical forms and remove isomorphic
 solutions by sorting these.

 {\scriptsize
 \begin{verbatim}
     genRamseyCG(S, T, N, Graphs) :-
         encode(ramsey(S,T,N), Matrix, Constraints),
         bCompile(Constraints,CNF),
         projectVariables(Matrix, Booleans),
         solveAll(CNF,Booleans,Assignments),
         decode(Assignments,Matrix,Graphs0),
         maplist(canonic_graph(N), Graphs0, Graphs1),
         sort(Graphs1, Graphs).
 \end{verbatim}
 }

 The predicate \verb!encode/3! is presented below. It first creates an
 $\mathtt{N\times N}$ adjacency \verb!Matrix! with Boolean variables
 representing the object of the search for a Ramsey(S,T;N) graph. It
 then imposes three sets of constraints: (1) the call to
 \verb!lex_star/2! constrains the rows of \verb!Matrix! to be
 pairwise lexicographically ordered. This implements the symmetry break
 described in~\cite{CodishMPS13}; (2) the first call to
 \verb!no_clique/4!  constrains the graph represented
 by \verb!Matrix! to contain no independent set of size \texttt{S}, and
 (3) the second call to \verb!no_clique/4!  constrains the graph
 represented by \verb!Matrix! to contain no clique of size \texttt{T}.
 The full details of the example are available for download as part of
 the \plnauty~library, in the \verb|examples| directory.

 {\scriptsize
 \begin{verbatim}
     encode(ramsey(S,T,N), map(Matrix), Constraints) :-
             adj_matrix_create(N, Matrix),
             lex_star(Matrix, Cs1-Cs2),          /* #1 */
             no_clique(0, S, Matrix, Cs2-Cs3),   /* #2 */
             no_clique(1, T, Matrix, Cs3-Cs4),   /* #3 */
             Cs4 = [],
             Constraints = Cs1.
 \end{verbatim}
 }

 The following illustrates the \bee\ constraint model, with the
 associated adjacency matrix, produced by a call to the
 \texttt{encode/3} predicate for a Ramsey $R(3,3;5)$ instance. Note
 that the elements on the diagonal of the matrix are $-1$ which is how
 \textit{false} is represented in \bee. The constraint model consists
 of three types of constraints corresponding to the three annotated calls
 in \verb!encode/3!.
 {\scriptsize
 \begin{verbatim}
   [[-1,A,B,C,D],
    [A,-1,E,F,G],
    [B,E,-1,H,I],
    [C,F,H,-1,J],
    [D,G,I,J,-1]]

     % #1 pairwise lexicographical order  % #2 no independent set    % #3 no clique
     bool_arrays_lex([B,C,D],[E,F,G]),    bool_array_or([A,B,E]),    bool_array_or([-A,-B,-E]),
     bool_arrays_lex([A,B,D],[F,H,J]),    bool_array_or([A,C,F]),    bool_array_or([-A,-C,-F]),
     bool_arrays_lex([A,F,G],[B,H,I]),    bool_array_or([A,D,G]),    bool_array_or([-A,-D,-G]),
     bool_arrays_lex([A,E,F],[D,I,J]),    bool_array_or([B,C,H]),    bool_array_or([-B,-C,-H]),
     bool_arrays_lex([B,E,I],[C,F,J]),    bool_array_or([B,D,I]),    bool_array_or([-B,-D,-I]),
     bool_arrays_lex([C,F,H],[D,G,I]),    bool_array_or([C,D,J]),    bool_array_or([-C,-D,-J]),
                                          bool_array_or([E,F,H]),    bool_array_or([-E,-F,-H]),
                                          bool_array_or([E,G,I]),    bool_array_or([-E,-G,-I]),
                                          bool_array_or([F,G,J]),    bool_array_or([-F,-G,-J]),
                                          bool_array_or([H,I,J]),    bool_array_or([-H,-I,-J]) 
 \end{verbatim}
 }

 Table~(\ref{tab:35n_sat}) summarizes the enumeration of all
 non-isomorphic $(3,5;N)$ colorings graphs using the constrain and
 generate approach. The first row indicates the number of
 (non-isomorphic) colorings generated. The second row indicates the
 number of colorings found when solving the constraint model (with the
 partial symmetry break).  The next rows detail the time (in seconds)
 to compute these colorings and the time spent in the calls to
 \verb!canonic_graph!.  It is notable that the time spent to reduce
 solutions modulo isomorphism using \nauty\ is negligible.

 \begin{table}[h]
 \centering 
 {\tiny
 \begin{tabular}{ c|cccccccccccccc}
   $n$  & 
   $1$  &     
   $2$  &    
   $3$  &   
   $4$  &  
   $5$  & 
   $6$  &
   $7$  &
   $8$  &
   $9$  &
   $10$ &
   $11$ &
   $12$ &
   $13$ &
   $14$ \\ 
 \hline
   $|\RR(3,5;n)|$  
   & 1          
   & 2             
   & 3             
   & 7             
   & 13            
   & 32            
   & 71            
   & 179           
   & 290           
   & 313           
   & 105           
   & 12            
   & 1             
   & 0  \\     
   $\#SAT$
   & 1
   & 2
   & 3
   & 7
   & 18
   & 63
   & 255
   & 1100
   & 3912
   & 7319
   & 3806
   & 272
   & 2
   & 0  \\
  time (sec)
  & 0.00 
  & 0.00 
  & 0.00 
  & 0.00 
  & 0.00 
  & 0.00 
  & 0.02 
  & 0.02 
  & 0.12 
  & 0.74 
  & 1.97 
  & 1.16 
  & 1.15 
  & 0.07 \\
  nauty (sec)
  & 0.00 
  & 0.00 
  & 0.00 
  & 0.00 
  & 0.00 
  & 0.00 
  & 0.00 
  & 0.01 
  & 0.05 
  & 0.16 
  & 0.05 
  & 0.00 
  & 0.00 
  & 0.00 
 \end{tabular}
 }
 \caption{Enumerating $R(3,5;n)$ graphs: Constrain, Generate \& Reduce.}
 \label{tab:35n_sat}
 \end{table}

 %

 \subsection{The \texttt{graph\char`_convert/5} predicate}

 The \texttt{graph\char`_convert/5} predicate performs conversions
 between the different graph formats that are supported by \plnauty.
 Supported formats include: adjacency matrices, adjacency lists, edge
 lists, and the \texttt{graph6} format.
 %
 As an example, to convert a graph, or a list of graphs, from the \texttt{graph6} format,
 to Prolog's adjacency matrix format:
 {\scriptsize
 \begin{verbatim}
      ?- Graph = `DqK', 
         graph_convert(5, graph6_atom, adj_matrix, Graph, AdjMatrix).

      Graph = `DqK',
      AdjMatrix = [[0,1,1,0,0], [1,0,0,1,0], [1,0,0,0,1], [0,1,0,0,1], [0,0,1,1,0]]
 \end{verbatim}
 }
 %
 {\scriptsize
 \begin{verbatim}
      ?- Graphs = [`DRo',`Dbg',`DdW',`DLo',`D[S',`DpS',`DYc',`DqK',`DMg',`DkK',`Dhc',`DUW'],
         maplist(graph_convert(5, graph6_atom, adj_matrix), Graphs, AdjMatrices).

      Graphs      = [`DRo',`Dbg',`DdW',`DLo',`D[S',`DpS',`DYc',`DqK',`DMg',`DkK',`Dhc',`DUW'],
      AdjMatrices = [[[0,0,1,0,1],[0,0,0,1,1],[1,0,0,1,0],[0,1,1,0,0],[1,1,0,0,0]],
                     [[0,1,0,0,1],[1,0,0,1,0],[0,0,0,1,1],[0,1,1,0,0],[1,0,1,0,0]],
                     [[0,1,0,1,0],[1,0,0,0,1],[0,0,0,1,1],[1,0,1,0,0],[0,1,1,0,0]],
                     [[0,0,0,1,1],[0,0,1,0,1],[0,1,0,1,0],[1,0,1,0,0],[1,1,0,0,0]],
                     [[0,0,1,1,0],[0,0,1,0,1],[1,1,0,0,0],[1,0,0,0,1],[0,1,0,1,0]],
                     [[0,1,1,0,0],[1,0,0,0,1],[1,0,0,1,0],[0,0,1,0,1],[0,1,0,1,0]],
                     [[0,0,1,0,1],[0,0,1,1,0],[1,1,0,0,0],[0,1,0,0,1],[1,0,0,1,0]],
                     [[0,1,1,0,0],[1,0,0,1,0],[1,0,0,0,1],[0,1,0,0,1],[0,0,1,1,0]],
                     [[0,0,0,1,1],[0,0,1,1,0],[0,1,0,0,1],[1,1,0,0,0],[1,0,1,0,0]],
                     [[0,1,0,1,0],[1,0,1,0,0],[0,1,0,0,1],[1,0,0,0,1],[0,0,1,1,0]],
                     [[0,1,0,0,1],[1,0,1,0,0],[0,1,0,1,0],[0,0,1,0,1],[1,0,0,1,0]],
                     [[0,0,1,1,0],[0,0,0,1,1],[1,0,0,0,1],[1,1,0,0,0],[0,1,1,0,0]]]
 \end{verbatim}
 }

 \subsection{The \texttt{canonic\_graph/6} predicate}

 The \texttt{canonic\_graph/6} predicate performs graph canonization
 and it takes the form \texttt{canonic\_graph(N, InputFmt, OutputFmt,
   Graph, Perm, Canonic)} where \texttt{InputFmt} is the format of the
 \texttt{N} vertex input graph (\texttt{Graph}), \texttt{OutputFmt} is
 the format of the canonical graph (\texttt{Canonic}), and \texttt{Perm}
 is the permutation whose application to the input graph renders the
 canonical representative.
 For example:
 {\scriptsize
 \begin{verbatim}
      ?- N = 5,
         Graph = [[0,1,0,0,0], [1,0,1,0,1], [0,1,0,1,0], [0,0,1,0,1], [0,1,0,1,0]],
         canonic_graph(N,adj_matrix,adj_matrix,Graph,Perm,Canonic).

      N       = 5,
      Graph   = [[0,1,0,0,0],[1,0,1,0,1],[0,1,0,1,0],[0,0,1,0,1],[0,1,0,1,0]],
      Canonic = [[0,0,0,0,1],[0,0,0,1,1],[0,0,0,1,1],[0,1,1,0,0],[1,1,1,0,0]],
      Perm    = [1, 5, 2, 4, 3]
 \end{verbatim}
 }

 A compact version of \texttt{canonic\_graph/6} is also included in
 \plnauty\ in the form of the predicate \texttt{canonic\char`_graph/3}. 
 The predicate \texttt{canonic\char`_graph/3} takes the form 
 \texttt{canonic\char`_graph(NVert, Graph, Canonic)} and it is 
 equivalent to \texttt{canonic\char`_graph(NVert, adj\char`_matrix,
 adj\char`_matrix, Graph, \char`_, Canonic)}.
 For example:
 {\scriptsize
 \begin{verbatim}
      ?- N = 5,
         Graph = [[0,1,0,0,0], [1,0,1,0,1], [0,1,0,1,0], [0,0,1,0,1], [0,1,0,1,0]],
         canonic_graph(N,Graph,Canonic).

      N       = 5,
      Graph   = [[0,1,0,0,0],[1,0,1,0,1],[0,1,0,1,0],[0,0,1,0,1],[0,1,0,1,0]],
      Canonic = [[0,0,0,0,1],[0,0,0,1,1],[0,0,0,1,1],[0,1,1,0,0],[1,1,1,0,0]]
 \end{verbatim}
 }

 \subsection{The \texttt{isomorphic\char`_graphs/6} predicate}

 The \texttt{isomorphic\char`_graphs/6} predicate tests for graph
 isomorphism. It takes the form: \texttt{isomorphic\char`_graphs(N,
   Graph1, Graph2, Perm, Canonic, Opts)} and tests if the two
 \texttt{N} vertex input graphs, \texttt{Graph1} and \texttt{Graph2}
 are isomorphic via a permutation \texttt{Perm}. If they are then
 \texttt{Canonic} is the canonical form they share. The predicate takes
 a list \texttt{Opts} of options to customize the behavior of this
 predicate.  Options  include any of the following:
 \texttt{fmt1(Fmt1)} the format of \texttt{Graph1},
 \texttt{fmt2(Fmt2)} the format of \texttt{Graph2},
 \texttt{cgfmt(CgFmt)} the format of \texttt{Canonic}.
 In the case where \texttt{Graph1} and \texttt{Graph2} are
 not isomorphic the predicate will fail silently. 
 %
 %
 For example:
 {\scriptsize
 \begin{verbatim}
      ?- N = 5,
         Graph1 = [[0,1,0,1,1], [1,0,1,0,0], [0,1,0,1,0], [1,0,1,0,1], [1,0,0,1,0]],
         Graph2 = [[0,1,0,1,1], [1,0,1,0,0], [0,1,0,0,1], [1,0,0,0,1], [1,0,1,1,0]],
         isomorphic_graphs(N, Graph1, Graph2, Perm, Canonic, []).

      N       = 5,
      Graph1  = [[0,1,0,1,1],[1,0,1,0,0],[0,1,0,1,0],[1,0,1,0,1],[1,0,0,1,0]],
      Graph2  = [[0,1,0,1,1],[1,0,1,0,0],[0,1,0,0,1],[1,0,0,0,1],[1,0,1,1,0]],
      Perm    = [1,2,3,5,4],
      Canonic = [[0,1,0,1,0],[1,0,0,0,1],[0,0,0,1,1],[1,0,1,0,1],[0,1,1,1,0]]

      ?- N = 5,
         Graph1 = [[0,1,1,0,1],[1,0,0,0,1],[1,0,0,0,0],[0,0,0,0,0],[1,1,0,0,0]],
         Graph2 = [[0,1,0,0,1],[1,0,1,1,0],[0,1,0,0,1],[0,1,0,0,1],[1,0,1,1,0]],
         isomorphic_graphs(N, Graph1, Graph2, Perm, Canonic, []).

      false.
 \end{verbatim}
 }

 \subsection{The \texttt{densenauty/8} predicate}

 Most of the core predicates of \plnauty\, and many of the examples 
 described above are based on the \texttt{densenauty/8} predicate. 
 The \texttt{densenauty/8} predicate is a direct interface to the 
 \nauty\ C library function of the same name. The predicate is called 
 in a similar fashion to its counterpart in the \nauty\ C library. A complete 
 documentation of \texttt{densenauty/8} may be found in the source code
 provided with \plnauty, and in the \nauty\ user guide \cite{nauty_guide}.

 Briefly, the predicate \texttt{densenauty/8} takes the following form:
 \begin{verbatim}
  densenauty(NVert, Graph, Labeling, Partition, 
             Permutation, Orbits, Canonic, Opts)
 \end{verbatim}
 where \texttt{NVert} is the number of vertices in the input graph, 
 \texttt{Graph} is the input graph, \texttt{Labeling}, \texttt{Partition}
 and \texttt{Orbits} are the labeling, partition and orbits of the 
 input graph, as described in the \nauty\ user guide \cite{nauty_guide},
 \texttt{Canonic} is the canonical form of the input graph, and 
 \texttt{Permutation} is the permutation of the nodes of the input graph
 which may be applied to obtain the Canonic representative. 
 The last argument, \texttt{Opts} is used to modify the behavior of 
 \texttt{densenauty}. For example, it may be used to control the format 
 of the input graph, and Canonic representative.



 \section{Interfacing Prolog and \gtools}
 \label{sec:plg}

 The \nauty\ graph automorphism tool comes with a collection of
 applications called \gtools, that implement an assortment of common
 tasks that \nauty\ is typically applied to. During installation (of
 \nauty) these are compiled into a set of command line applications.
 These applications cannot simply be loaded using the foreign language
 interface. Each application is like a black box. We do not wish to
 access its source code. 
 One straightforward approach to integrate  \gtools\ with Prolog is to
 run each such application from within Prolog, write its output to a
 temporary file, and then to  read the file, and continue with the task
 that the Prolog program is addressing.

 A more elegant solution makes use of unix pipes to skip that
 intermediate step of writing and reading from files. The output is
 directly read/written via Prolog. The voodoo is using pipes (which are
 like in-memory files).  We have implemented a Prolog library called
 \plgtools, which provides a framework for calling the applications in
 \gtools\ using unix pipes.
 The \plgtools\ framework supports two types of \gtools\ applications
 which take any number of command line arguments and write their output
 to standard output. The first type does not require any input, and the
 second  requires some form of input (from standard input).
 We present a general template to support the two ``sides'' of the
 pipe: a child predicate (which typically executes a \gtools\ command),
 and a parent predicate (which typically reads the output of the
 child).



 The  framework includes  predicates:
 \verb!gtools_exec/6! and \verb!gtools_fetch/2!, and
 two additional predicates for applications which respectively require
 uni- and bi-directional communication: \verb!gtools_fork_exec/2!  and
 \verb!gtools_fork_exec_bidi/2!.
 For uni-directional communication, a call to
 \verb!gtools_fork_exec(Parent, Child)! will fork and execute the
 \texttt{Parent} goal as the parent process and the \texttt{Child} goal
 as the child process. It assumes that both \texttt{Parent}
 and \texttt{Child} take an additional argument which is unified with
 the corresponding input/output streams (to support communication from
 child to parent).
 For bi-directional communication, a call to
 \verb!gtools_fork_exec_bidi(Parent, Child)! is exactly the same,
 except that the \texttt{Parent} and \texttt{Child} take two additional
 arguments to support two way communication.

 The predicate \verb!gtools_fetch/2!  reads the next line from the
 output stream of the child and converts it to an atom. When the end of
 the stream is reached then the predicate fails.
 A call to \verb!gtools_exec/6! takes the form 
 {\small 
 \texttt{gtools\_exec(NautyDir, Cmd, Args, InputStream, OutputStream, ErrorStream)}
 }
 where: \verb!NautyDir! is the directory in the file system which
 contains the \gtools\ applications, \verb!Cmd! is the name of the
 \gtools\ command that we which to execute, and \verb!Args! is its
 argument list. The final three arguments specify the standard input,
 output and error streams. 
 The call to \verb!gtools_exec/6! invokes the \texttt{exec/1} predicate
 of SWI-Prolog, replacing the current process image with \texttt{Cmd}
 and its \texttt{Args}.

 We present two example uses of \plgtools. The first, calls
 \texttt{geng} from \gtools, which iterates over all non-isomorphic
 graphs with a given number of vertices. The second,  calls
 \texttt{shortg} from \gtools, which reduces a set of graphs to
 non-isomorphic members.

 \subsection{Example 1: \texttt{geng}}
 \label{sec:geng}

 This example illustrates how the framework is applied for an
 application which reads no input. 
 The \gtools\ application \texttt{geng} receives an argument
 \texttt{n} and outputs one line for each non-isomorphic graph with $n$
 vertices.
 Its Prolog implementation consists of three predicates:
 \texttt{geng/2}, \texttt{parent\_geng/2} and
 \texttt{child\_geng/2}. The predicate \texttt{geng/2} is the main
 predicate which backtracks over all results of the \gtools\
 application. The predicates \texttt{parent\_geng/2} and
 \texttt{child\_geng/2} implement respectively the parent and child
 sides of the pipe. 
 {\scriptsize
 \begin{verbatim}
      geng(N, Graph) :-
             gtools_fork_exec(geng:parent_geng(Graph), geng:child_geng(N)).

      parent_geng(Graph,Read) :-
             gtools_fetch(Read, Graph).

      child_geng(N,Stream) :-
             gtools_exec(`nauty26r3', geng, [`-q', N], _, Stream, _).
     \end{verbatim}
 }
 \subsection{Example 2: \texttt{shortg}}

 This example illustrates how the framework is applied for an
 application which reads from standard input. 
 The \texttt{shortg} application reads a list of graphs in the
 \texttt{graph6} format~\cite{graph6} from standard input, and removes all isomorphic
 duplicates, writing to standard output. It can be applied as follows:

 After integrating \texttt{shortg} with \plgtools\ it could be 
 called from Prolog like so:
 {\scriptsize
 \begin{verbatim}
      ?- InputGraphs = [`DRo',`Dbg',`DdW',
                        `DLo',`D[S',`DpS',
                        `DYc',`DqK',`DMg',
                        `DkK',`Dhc',`DUW'], % a list of graphs in graph6 format
         shortg(InputGraphs, OutputGraphs). % the call to shortg

      InputGraphs  = [`DRo',`Dbg',`DdW',`DLo',`D[S',`DpS', `DYc',`DqK' | ... ],
      OutputGraphs = [`DqK'].
 \end{verbatim}
 }

 The implementation of \texttt{shortg} in Prolog consists of
 three predicates and is very similar to that for \texttt{geng}
 except that communication between the child and parent processes is
 bi-directional. 

 {\scriptsize
 \begin{verbatim}
     shortg(In, Out) :-
          gtools_fork_exec_bidi(shortg:parent_shortg(In, Out), shortg:child_shortg).

     parent_shortg(In, Out, PRead, PWrite) :-
             maplist(writeln(PWrite), In),
             flush_output(PWrite),
             close(PWrite),
             findall(O, gtools_fetch(PRead, O), Out),
             close(PRead).

     child_shortg(CRead, CWrite) :-
             gtools_exec(`nauty26r3', shortg, [`-q'], CRead, CWrite, _).
 \end{verbatim}
 }
 In this example,  \texttt{shortg/2} takes two arguments:
 \texttt{In} a list of input graphs in the \texttt{graph6} format,
 to be reduced modulo isomorphism, and 
 \texttt{Out} will be unified with the set of reduced
 graphs. The predicate calls the \texttt{gtools\_fork\_exec\_bidi/2}
 predicate. Pipes are opened to setup two way communication between the
 parent and child.

 Two additional predicates are implemented: one for the parent process and
 one for the child process. Each predicate takes, as its last two arguments
 the read and write ends of the pipes, so communication may be established.
 In our case, the parent writes the set of input graphs to the write
 end of the pipe, and then reads the results from the read end of the 
 child's pipe. The child calls \texttt{gtools\_exec/6}, and executes
 \texttt{shortg/2}.

 %
 %

 \section{Technical Details}
 \label{sec:tech}

 A short overview of some technical details regarding \plnauty\ and 
 \plgtools\ follows. 

 The package containing \plnauty\ and \plgtools\ is available for download
 from the \plnauty\ homepage at: \url{http://www.cs.bgu.ac.il/~frankm/plnauty}.
 The package contains a \texttt{README} file, which contains usage and 
 installation instructions, as well as an \texttt{examples} directory 
 containing the examples discussed in this paper. The C code for \plnauty\ may
 be found in the \texttt{src} directory. Also in the \texttt{src} directory
 are the two module files for \plnauty\ and \plgtools.

 Both \plnauty\ and \plgtools\ were compiled and 
 tested on Debian Linux and Ubuntu Linux using the 7.x.x branch of 
 SWI-Prolog. It is important to mention that both \plnauty\ and 
 \plgtools\ contain Linux specific features, and are oriented towards 
 SWI-Prolog. It should also be noted that \plnauty\ is not thread-safe, 
 for reasons of performance. If you require a thread-safe version of 
 \plnauty\ you should synchronize calls to the predicates of the 
 \plnauty\ module.



 %


 \section{Conclusion}
 \label{sec:conclude}

 We have presented, and made available, a Prolog interface to the core
 components of the \nauty~graph-automorphism tool~\cite{nauty} which is
 often cited as ``The world's fastest isomorphism testing program''
 (see for example
 {\small\url{http://www3.cs.stonybrook.edu/~algorith/implement/nauty/implement.shtml}}).
 The contribution of the paper is in the utility of the tool which we
 expect to be widely used. The tool facilitates programming with the
 strengths of two paradigms: logic programming for solving graph search
 problems on the one hand, and efficient pruning of (intermediate)
 solutions modulo graph isomorphism, on the other.
 It enables Prolog programs which address graph search problems to
 apply \nauty~natively, through Prolog, in the process of graph search
 and enumeration. Graphs may be generated non-deterministically and may
 be canonized deterministically.
 %




\end{document}